\documentclass[12pt, dvipsnames]{article}

\usepackage{verbatim, booktabs, multirow, hyperref}
\usepackage{graphicx}
\usepackage[round]{natbib}
\usepackage[utf8]{inputenc}
\usepackage{enumerate}
\usepackage{euscript}
\usepackage{amsmath, amsthm, amsfonts, amssymb}
\usepackage[margin = 1in]{geometry}
\usepackage{pgfplots,tikz}
\usepackage{bm}
\usepackage{pifont}
\usepackage{bm}
\usetikzlibrary{calc} 
\usetikzlibrary{positioning}
\usepackage{lscape} 
\usepackage{subcaption}
\usepackage{framed}

\usepackage{multicol,multirow,booktabs}

\usepackage{algorithm}
\usepackage{algpseudocode}

\usepackage[normalem]{ulem}

\usepackage[title]{appendix}

\usetikzlibrary{arrows.meta}
\usepackage{apptools}

\theoremstyle{plain}
\newtheorem{thm}{Theorem}

\newtheorem{cor}{Corollary}

\theoremstyle{definition}
\newtheorem{defn}{Definition}

\newtheorem{remark}{Remark}

\newtheorem*{example*}{Example}

\newtheoremstyle{named}{}{}{\itshape}{}{\bfseries}{.}{.5em}{\thmnote{#3} #1}
\theoremstyle{named}
\newtheorem{rpproperty}{}

\newtheoremstyle{named}{}{}{\itshape}{}{\bfseries}{.}{.5em}{#1 \thmnote{#3}}
\theoremstyle{named}

\AtAppendix{\counterwithin{lemma}{section}}
\AtAppendix{\counterwithin{prop}{section}}
\AtAppendix{\counterwithin{thm}{section}}

\newcommand{\argmin}{\operatornamewithlimits{argmin}}

\newcommand{\cF}{\mathcal{F}}

\newcommand{\cM}{\mathcal{M}}

\newcommand{\gB}{\mathcal{B}}

\newcommand{\piv}{\texttt{piv}}

\newcommand{\ds}{(\gB,\C)}
\newcommand{\dsi}{(\gB_1,\C_1)}
\newcommand{\dsii}{(\gB_2,\C_2)}

\newcommand{\cha}{\C(\gB)}

\let\emptyset\varnothing

\title{On the Welfare (Ir)Relevance of Two-Stage Models\thanks{
We would like to thank Yuta Inoue and Koji Shirai for providing the data used in this paper.}
}
\author{Mikhail Freer\thanks{Department of Economics, University of Essex.  e-mail: \texttt{m.freer@essex.ac.uk}} \and Hassan Nosratabadi\thanks{ECARES, Universit\'e Libre de Bruxelles. e-mail: \texttt{seyed.nosratabadi@ulb.be}}}

\newcommand{\C}{\bm{c}}

\begin{document}

\maketitle

\begin{abstract}
    In a two-stage model of choice a decision maker first shortlists a given menu and then applies her preferences.
    We show that a sizable class of these models run into significant issues in terms of identification of preferences (welfare-relevance) and thus cannot be used for welfare analysis.
    We classify these models by their revealed preference principles and expose the principle that we deem to be the root of their identification issue.
    Taking our analysis to an experimental data, we observe that half of the alternatives that are revealed preferred to another under rational choice are left revealed preferred to nothing for any member of this class of models.
    Furthermore, the welfare-relevance of the specific models established in the literature are much worse.
    The model with the highest welfare-relevance produces a revealed preference relation with the average density of 2\% (1 out of 45 possible comparisons revealed), while rational choice does 63\% (28 out of 45 possible comparisons).
    We argue that the issue is not an inherent feature of two-stage models, and rather lies in the approach with which the first stage is modeled in the literature. 

\end{abstract}

\section{Introduction}

Models of bounded rationality have long been of interest as a way to accommodate ``irrational'' patterns.
The standard approach of these models is to generalize rational choice model by adding degrees of freedom; i.e., new primitives of the model.
Examples are adding shortlisting stages, reference points or perturbing the preference relations.
However, enriching the set of primitives inevitably leads to potential identification problems due to the increased degrees of freedom.
Intuitively then one could conjecture that such models would be weak in terms of identification power compared to rational choice.
Recent papers in the literature  (see \cite{dardanoni2020inferring} and \cite{dardanoni2023mixture}) acknowledge this and hence suggest considering population data instead, hinting to a drawback that such models impose a ``richness'' condition on the data ``comprising a single individual’s choices from a \textit{large number} of different \textit{overlapping} menus''. 

 

Two-stage models constitute a large class of models of bounded rationality.
In these models a decision maker (DM) first makes a shortlist from a given problem and then uses her preferences to choose the best alternative from the shortlisted ones.
One particularly interesting observation that speaks to the identification problem in many of these models is that if the data set is consistent with rational choice, then one cannot identify preferences \emph{at all}.\footnote{This for example is the case for the models in \cite{masatlioglu2012revealed}, \cite{lleras2017more}, \cite{manzini2007sequentially}, and \cite{au2011sequentially} among many others.}
This ``inconsistency-dependency'' is a rather paradoxical feature in that the models that are aimed to generalize rational choice can only be identified if observed choice is not rational.

In this paper, we follow the revealed preference tradition to better understand the issue of preference identification in two-stage models.
This tradition speaks to establishing a proper mapping between the set of observables (choice) to the set of unobservables of the decision process. (preferences, cognitive constrains, etc.)
Thus we try to find the common observational requirements these models impose to reveal preferences.
we refer to these requirements as \emph{revealed preference principles} (RP principles).
In particular, we define these principles in terms of an alternative being revealed preferred to at least another (having a non-empty lower contour set).

The first principle we introduce is a stringent one called \emph{choice over pivots} (\textbf{CoP}).
A \emph{pivot} is an alternative the removal of which alters choice.
\textbf{CoP} implies that an alternative that is not at least chosen over a pivot in a data point, will have an empty lower contour set.
For rational choice, in contrast, each choice observation is welfare-relevant: the chosen alternative is better than those that are available but not chosen.
Four prominent two-stage models follow \textbf{CoP}.
These models are (i) \emph{limited attention} (\cite{masatlioglu2012revealed}), (ii) \emph{limited consideration} (\cite{lleras2017more}), (iii) \emph{rational shortlist method} (\cite{manzini2007sequentially}), and (iv) \emph{transitive rational shortlist method} (\cite{au2011sequentially}).

The second RP principle is \textit{standard transitive closure} (\textbf{STC}): if a data set does not help identifying any information about the lower contour set of an alternative, then adding more data where the alternative is not available at all can not help with the identification.
That is, as in rational choice, the indirect revealed preference relation can only built upon an already non-empty direct revelation.
This is a rather uncontroversial principle that intuitively should hold for almost any standard two-stage model.
Indeed, in addition to the four models above, rational choice follows it as well.
It turns out that, when combined with \textbf{CoP}, this new property exacerbates the observational burden on welfare analysis: an alternative now needs to be \textit{involved in a violation of the weak axiom of revealed preferences} (WARP) to be have a non-empty lower contour set, where involvement means \textit{being chosen in the intersection of a WARP violation}.\footnote{Many other two-stage models do run into this welfare-irrelevance result. Examples are \cite{lleras2021path}, \cite{inoue2018limited}, \cite{yildiz2016list}, \cite{kimya2018choice}, \cite{geng2021shortlisting}, \cite{horan2016simple}, and \cite{geng2022limited}. For the sake of being concise we do not include them in our analysis.}

We next introduce an even more stringent RP principle: \emph{choice over choice} (\textbf{CoC}): an alternative that is not chosen over a chosen alternative has an empty lower contour set.
To see why this property is more stringent than \textbf{CoP}, note that there are two type of pivots : chosen points and unchosen points.
\textbf{CoC} immediately disregards the latter type of pivots for revealed preference purposes.
All models except (iv) follow \textbf{CoC}.
When combined with \textbf{STC}, the extra observational burden becomes the following: for an alternative to have a non-empty lower contour set it has to be \textit{directly} involved in a WARP violation -- i.e., cause the WARP violation itself.\footnote{
    Note that neither \textbf{CoP} nor \textbf{CoC} implies irrational patters in choice. It is indeed combining it with \textbf{STC} that makes preference identification dependent on WARP violations. For a discussion, see section \ref{sec:wir}.}


The important point about our result is that the inconsistency-dependency feature is \textit{not an innate} feature of a two-stage model, and rather due to the formulation of a typical two-stage model in the literature.
Take for example a model where DM shortlists with mechanics in the first stage unknown to the observer, but in a way that at least two alternatives are considered in the second stage.
The intuition of this assumption is that DM faces a trade-off in the second stage.
In such a model, welfare analysis can be done \textit{whether or not} the data exhibits irrational patters.
Since we argue that \textbf{STC} is a a rather non-restrictive assumption, the underlying reason for identification issues are \textbf{CoP} and \textbf{CoC}.
Indeed, a model with this size-dependent formulation of the first stage still satisfies \textbf{STC}, and thus moves beyond the inconsistency-dependent feature precisely because it violates \textbf{CoP}.
Finally, our results also indicate that \textit{not all} patterns of violations are welfare-relevant in these models.
To be more precise, the violation of \textit{strong axiom of revealed preferences} (SARP) are yet welfare-irrelevant if they are not WARP violations.\footnote{That is, choice cycles of orders higher than two. It is worth noting that the dissection between the two types of violations is only possible if the data set is incomplete. For a discussion on this see Section \ref{sec:wir}.}
 
 
We complement our analysis by an experimental illustration.
We first report welfare-irrelevance results regarding our theoretical analysis based on \textbf{CoP(C)}.
We show that half of the alternatives that have non-empty lower contour sets under rational choice will be left with an empty lower contour set in \textit{any} two-stage model that follows \textbf{CoP(C)} and \textbf{STC}. 
After providing the general analysis of the class of models, we then zoom in and provide the welfare-relevance of the specific two-stage models (i)-(iv).
Using this specific models allows us to provide precise results on welfare-relevance by talking about the density of the revealed preference relation.
That is, the number of comparisons revealed given a model normalized by the total number of comparisons existing in a complete and acyclic binary relation.
This precise welfare-relevance measure is drastically lower for these specific models when compared to the general drop observed in the number of alternatives with non-empty lower contour sets.
The best model is (iv) (the only non-\textbf{CoC} model in the list above) for which only 2\% of preferences are revealed, while for rational choice 63\% of comparisons are revealed.
These results illustrate that, even in the class of two-stage models that follow \textbf{CoP} and \textbf{STC}, these specific models deliver minimal welfare-relevance.

Next, we investigate versions of these model where we force a \textbf{CoP}-violating assumption in order to see whether or not we can improve their welfare-relevant. For this objective, we consider amended versions of these models by assuming that DM (in addition to following the rules of the model) also always considers at least two alternatives.
The amended versions, as discussed before, violate \textbf{CoP} and thus does not have to follow the inconsistency-dependent results.
These models deliver two important results.
First, the average density of revealed preferences relation rises to about 20\% which is about a third of density for the rational choice.
This is a rather significant increase and potentially sufficient in many applications given that we are dealing with two-stage models with an extra primitive. 
More importantly, this rise is almost the same regardless of the model (densities are in the range 17\% to 21\%), hinting that the \textbf{CoP}-violating assumption (the size dependent formation) is by far more relevant for preference identification than the conditions imposed on the first stage of these model.



Our analysis assumes that the observer only has access to an incomplete (realistic) data set. The analysis of two-stage models given such data set has received attention recently.
\cite{de2021bounded} investigates the question of \textit{falsifiability} of a variety of behavioral models (some of which are in our analysis as well) in such a setting. They provide a common way for testing the models; that is to see if an observed data set could be produced by a given model. Our question is instead on \textit{identifiability} of preferences within these models. 

As mentioned earlier, a simple way to break the inconsistency-dependent identification is to ensure that DM performs some comparisons after shortlisting.
This idea has been used in the literature before, for example in \cite{barseghyan2021discrete} and \cite{bajraj2015choosing}.\footnote{The latter paper in particular introduces a two-stage model where DM picks the top two alternative based on a linear order in the first stage, and then applies another linear order (preference relation) in the second stage.
Thus this model is slightly different from the size-dependent formulation we mentioned earlier because there more than two alternative may appear in the second stage.}
In addition, the consideration capacity model in \cite{dardanoni2020inferring} is reminiscent of the same principle.

The remainder of this paper is organized as follows.
Section 2 presents necessary definitions.
Section 3 presents the theoretical results.
Section 4 provides experimental illustration.
Section 5 concludes.
All proofs omitted in the text can be found in the Appendix.

\section{Definitions}

Let $X$ be a finite universal set of alternatives.
Let $\gB\subseteq 2^X$ be a collection of choice problems with at least two alternatives ($|B| \geq 2$ for all $B \in \gB$) where $2^X$ is the set of all the subsets of $X$.
$\C:\gB\rightarrow X$ is a {\bf choice function} if $\C(B)\in B$.
Thus a {\bf data set} can be described by the pair $(\gB,\C)$.
Let $\C(\gB) = \bigcup_{B\in \gB} \C(B)$ be the collection of all \textbf{chosen points} in the data set.
A binary relation $\succ\; \subseteq X\times X$ is a strict preference relation if it is complete (any pair of alternatives is comparable), transitive ($x\succ x'$ and $x\succ x'$ implies $x\succ x''$ for any $x,x',x''\in X$), and asymmetric ($x\succ x'$ implies $x'\not\succ x$ for any $x,x'\in X$).
By $\max(B,\succ)$ we mean the set of \textbf{maximal} elements of $\succ$ in $B$; that is, $\max(B,\succ) = \{x\in B: \not\exists \;y \in B: y \succ x\}$.

\paragraph{Two-Stage Models}
In the first stage DM shortlists a given problem using a \textbf{filter}: a mapping $F: 2^X \rightarrow 2^X$ such that $F(B) \subseteq B$. 
We denote the set of all possible filters by $\mathfrak{F}$.
$F(B)$ thus may be understood as the \textit{consideration set} for the problem $B$.
The second stage, without loss of generality, could be assumed to consist of applying a well-defined preference relation on what survives the first stage.\footnote{An elaboration may be needed here. In general one may impose a weaker requirement on the binary relation in the second stage. For example, \cite{manzini2007sequentially} only requires an asymmetric binary relation in the second stage. However, in order for the choice function to be well-defined (i.e., non-empty valued) one need to at the end impose acyclicity over the items that survive the first stage. Acyclicity of course guarantees the existence of an extension to a strict linear order. This would mean that any two-stage model generates a well-defined choice function if and only if it makes uses of a strict linear order in the second stage. That is assuming second stage rationality is with no restriction on the explanatory power of the model. For a similar classification of two-stage models see \cite{tyson2013behavioral}.} 
Naturally we assume that, a priori, \textit{any} preferences in the second stage are allowed for any $\cF$. 
Consequently, the only distinguishing aspect of any given pair of models is their first stage; i.e., \emph{the collection of filters} they allow. 
A two-stage model $\mathcal{F}$ thus is simply a subset of $\mathfrak{F}$.
Rationalizability would then mean that one could find a filter in $\cF$ and a preference relation to generate the observed data with.

\begin{defn}
\label{def:tsm}
A \textbf{two-stage model} of choice is a collection of filters; i.e., $\cF$ is a two-stage model if and only if $\cF \subseteq \mathfrak{F}$. A data set $(\gB,\C)$ is \textbf{rationalizable} with model $\mathcal{F}$ if there exist
\begin{itemize}
    \item[(i)] a filter $F\in \mathcal{F}$, and
    \item[(ii)] a preference relation $\succ$,

    such that
$$
\C(B) = \max(F(B),\succ)
$$
for every $B\in \gB$.\footnote{Note that by definition a filter is a mapping defined on $2^X$. Thus, our definition of rationalizability entails the existence of an \textit{extension} of any incomplete data set to a complete one consistent with the model. To see the importance of the existence of such an extension, and the fact that is not guaranteed, see \cite{de2021bounded}.}
\end{itemize}

\end{defn}

The definition above is as general as possible. 
One can think of an extremely exclusive model such as $\cF_1$ that only includes one filter, say the identity filter: $F(B) = B$ for all $B$; i.e., the classical \emph{rational choice}.
$\cF_2$ could be another model that filters anything but $x_1$ out when $x_1$ is available, otherwise anything but $x_2$ is filtered out when $x_2$ is available, \ldots, otherwise anything but $x_n$ when $x_n$ is available, and where $X = \{x_1,x_2,\ldots,x_n\}$.
Of course one can conceive a model $\cF_3 = \mathfrak{F}$; i.e., a model with no restriction on filter selection. 
At the same time, $\cF_4 = \cF_1 \cup \cF_2$ is another model in this framework.

\paragraph{Revealed Preferences (RP)}

Let us next formalize the revealed preference relation. 
We need to introduce some additional notations first. 
Let
$$
\cM(\gB,\C) = \{(\succ,F) \text{ that rationalize } (\gB,\C) \}
$$
be the collection of all \textbf{the instances of the model that rationalize the data} set. 
For a rationalizable data set (i.e., $\cM\ds \neq \emptyset$), we can define the revealed preference relation as 
$$
\succ^* = \bigcap\limits_{(\succ,F)\in \cM(\gB,\C)} \succ.
$$


Also let $L(x|\ds)$ be the the set of alternatives revealed inferior to $x$ given the data set $\ds$ according to $\succ^*$; that is, the \textbf{lower contour set} of $x$. 

The last notation is the following. 
Consider a two-stage model where DM makes at least one comparison in the second stage; that is, the filters of size of at least two. 
When we observe the data point $\dsi = \C\{x,y,z\} = x$ what can we infer about preferences? 
We would know that $x \succ^* y$ \textbf{or} $x \succ^*z$.
These latter two pieces of information impose a ``restriction'' on the lower contour set of $x$, absent of extra observation, they nonetheless do not enrich RP; that is, $L(x|\ds) = \emptyset$.
The restrictions could be insightful with more data though. 
For example, if we collected another data point in the form of $\C\{x,z\} = z$, then we would know that $z \succ^* x$, and using the or-logic that $x \succ^* y$. 
This indicates the important of the or-logic in performing RP in two-stage models. 
In order to contain these information in our analysis then we use the notation $\mathcal{R}^{\downarrow} (x|\ds)$ to be the \textbf{set of all restrictions} on the lower contour set of $x$. 
In our example, we have $\mathcal{R}^{\downarrow} (x|\dsi) = \{(x \succ^* y \; \textbf{or} \; x \succ^* z$)\}.\footnote{Note that the $x$ need not be involved in all \textbf{or} segments, or on the same side to be a valid restriction. For example, ($x \succ^* y$ \textbf{or} $z \succ^* t$) and ($x \succ^* y$ \textbf{or} $z \succ^* x$) are also valid restrictions for $x$ and thus belong to $\mathcal{R}^{\downarrow} (x|\dsi)$. A second note is that \textbf{or}-logic could also appear in some of the prominent two-stage models in the literature. See \cite{de2021bounded} for a comprehensive discussion on this matter.}\footnote{Let us also mention that rigor requires us to, in addition to the data set, include the model from the lens of which we are inferring preferences as an argument in these notations; that is, to instead write: $\succ^*(\ds, \cF)$, $L(x | \ds, \cF)$, and $\mathcal{R}^{\downarrow} (x|\dsi,\cF)$. Unless needed though, we omit these extra arguments in order to keep the flow of the writing.}

\subsection{Prominent Two-Stage Models}\label{sec:models}


\begin{itemize}
    \item [--] \textbf{Limited Attention} \citep[LA, see][]{masatlioglu2012revealed}
    
    \textit{Attention Filters}: Filters are insensitive to removal of filtered alternatives.
    $$
    F\in \cF \text{ if and only if } F(B) = F(B\setminus\{x\}) \text{ if } x\notin F(B).
    $$ 
    
    \item [--] \textbf{Limited Consideration} \citep[LC, see][]{lleras2017more}
    
    \textit{Competition Filters}: The set of filtered alternatives may only shrink if the choice problem expands:
    $$
    F\in \cF \text{ if and only if } F(B) \cap B' \subseteq F(B') \text{ if } B'\subseteq B. 
    $$

        
    
   \end{itemize}


  \begin{itemize}
    \item [--] \textbf{Rational Shortlist Method} \citep[RSM, see][]{manzini2007sequentially}
    
    \textit{Asymmetric Binary Filter}: Irrespective of the choice problem, if $x$ filters $y$ out, then $y$ does not filter $x$ out. Formally, there is an asymmetric binary relation $\succ_0$ the maximal elements of which survive the first stage.
    $$
    F\in \cF \text{ if and only if } F(B) = \max(B,\succ_0). 
    $$

    \item [--]  \textbf{Transitive Rational Shortlist Method} \citep[TRSM, see][]{au2011sequentially}
    
    \textit{Asymmetric and Transitive Binary Filter}: In addition to asymmetry, and again irrespective of the choice problem, if $x$ filters $y$ out and $y$ filters $z$ out, then $x$ filters $z$ out. Formally, there is a asymmetric and transitive binary relation  $\succ_0$ that characterizes the filter:
    $$
    F\in \cF \text{ if and only if } F(B) = \max(\succ_0,B). 
    $$

\end{itemize}

\noindent
Note that, RSM is nested in LC (but not in LA). TRSM is of course nested in RSM and thus in LC, but it has been shown that it is also in both LA. These relationships are interesting to consider later where we quantify the welfare-relevance of these models.\footnote{
\label{fn:othermodels}It is important to note that while we concentrate on these four models due to their relevance in the literature, there are many other two-stage models that lie somewhere in the union of LA and LC. These models include but no restricted to \cite{lleras2021path}, \cite{cherepanov2013rationalization}, \cite{yildiz2016list}, \cite{kimya2018choice}, \cite{geng2021shortlisting}, \cite{horan2016simple}, and \cite{geng2022limited}. Our analysis in the next section also implies to these models.
}


\section{Welfare (Ir)Relevance}\label{sec:wir}

In this section we provide a set of fundamentals of two-stage models that should hint at their unimpressive identifiability (of preferences).
These fundamentals will speak to the revealed preference principles of a model -- i.e. the observational requirements for preference identification -- instead of the ``mechanics'' of its filters.\footnote{Focusing on RP principles instead of model properties is best understood by the works in \cite{bernheim2009beyond} (and later \cite{nishimura2018transitive}) where the authors propose focuses on RP principles that ``make sense'' abstract of a model (i.e., a model-free approach) as opposed to the decision theoretic properties that produce them (model-based approach). For a comprehensive discussion on the distinction between the two approaches see \cite{manzini2014welfare}.}
Let us start with some well-known concepts. Let $\ds$ be a data set.  
\begin{enumerate} 
    \item[i.] A \textbf{WARP} violation is a pair $(B_1,B_2): B_i \in \gB$ such that $\C(B_1) \neq \C(B_2)$, $\C(B_1),\C(B_2) \in B_1 \cap B_2$, and

    \item[ii.] a \textbf{SARP} violation is a tuple $(B_1,B_2,\ldots,B_n): B_i\in \gB$ such that $\C(B_i)\neq \C(B_j)$ for any $i\neq j$, $\C(B_i)\in B_{i+1}$ for every $i\le n-1$ and $\C(B_n)\in B_1$, and

    \item[iii.] a \textbf{pure SARP} violation is a SARP violation that is not a WARP violation. 
\end{enumerate}

\noindent
Next, we introduce a fundamental notion that repeatedly appears in the literature of two-stage models and plays a critical role in our analysis; i.e., \textit{pivots}.

\begin{defn}
    We call $p\in X$ a \textbf{pivot} in $\ds$ if there exists $S\in X$ such that $\C(S) \neq \C(S-p)$ in all the choice functions generated by all the rationalization of $\ds$. We denote the set of all pivot in the data set $\ds$ by $\piv(\gB,\C)$.
\end{defn}


    

\noindent
Hence, a pivot is an alternative removal of which causes a change in DM's choice. 
Note that if a data set is consistent with rational choice model then the only pivots are chosen alternatives; that is, $\cha$.\footnote{
This is immediate consequence of the fact that rational choice model satisfies \textit{independence of irrelevant alternatives}.}
We refer to this kind of pivot as \textbf{Type 1}.
If rationality is violated, we could have unchosen alternatives that are pivots. 
For example, take the choice pattern $\C\{x,y\} = y$ and $\C \{x,y,z\} = x$ where removing $z$ changes DM's choice. We refer to such pivots as \textbf{Type 2}.\footnote{
\cite{bajraj2015choosing} refers to pivots as ``reversers'' and uses a similar categorization of such elements (using the terms ``trivial'' and ``non-trivial'', respectively instead) in a model where exactly two alternatives survives the first stage.
}

\subsection{Choice over Pivots}

The key insight into the importance of pivots is the following: the removal of a pivot alters the consideration set \textit{irrespective} of the two-stage model (that is, for all $\cF$ in Definition \ref{def:tsm}).
The reason is that the different choice made in the new set (after removal) and the original set can not be rationalized with the same consideration set -- that is, if removal of $t$ alters choice in $B$, then $F(B) \neq F(B-t)$.
However, if the removal of an alternative does \textit{not} alter the choice in a given problem, then a construction where the consideration set before and after is unchanged seems both straightforward and intuitive.
This particular rationalization of course could only be the case if the removed alternative was not considered before the change; that is, alternatives that are not pivots may be assumed out of consideration sets. 
The latter assertion is indeed the link that connects various two-stage models. 
This common link, in turn, imposes a necessary condition on the revealed preference relation. 
The standard revealed preference approach maps the set of unobservables (preferences along with other decision elements) to the observed choice patterns. 
It is natural in such an approach that for an alternative that is never chosen to be not revealed better than any other.\footnote{This is standard in the sense that it also holds true under rational choice.}
On the other hand, the argument above tells us that if an alternative is considered is some choice problem that it is a pivot.
Consequently, a necessary condition for an alternative to be revealed superior to another is for it to be \textit{chosen over a pivot}. This principle is formalized next.

\begin{rpproperty}[Choice over Pivots (CoP)]
\label{axiom:Direct} 
If $x$ is not chosen over a pivot in a data set, then there are no restrictions on the lower contour set of $x$ given that data set; that is,\\

\noindent $\forall \ds, \forall x$, if $\not\exists y\in \piv(\gB,\C)$ and $B \in \gB$ such that $x= \C(B)$ and $y\in B$, then $\mathcal{R}^{\downarrow}(x | \ds) = \emptyset$. 

\end{rpproperty}

Before proceeding further let us make two important remarks about \textbf{CoP}.
First, this RP principle is quite demanding than compared to the RP principle under rational choice.
For an alternative to have non-empty lower contour set it is (necessary and) sufficient for this alternative to be \emph{chosen} in some problem.
Such principle is indeed not compatible with \textbf{CoP}.
In the two-stage world in which these models operate, to infer that $x$ is better than ``something'', it is \textit{necessary} to observe two patterns: (i) a data point that shows $x$ is chosen where $y$ is available, and (ii) a pair of data points that indicates $y$ is a pivot. 
Moreover, \textbf{CoP} is not a sufficient condition for RP. 
The second point is that one can easily contemplate a reasonable two-stage models that violates \textbf{CoP}. 
Assume that DM always has at least two alternatives in her consideration set. This model obviously violates \textbf{CoP} since any observation on a doubleton (such as $\C\{x,y\} = x$) in isolation allows us to infer DM's preferences over the pair. ($x \succ y$) 

While \textbf{CoP} is rather demanding, the second RP principle that we propose is instead uncontroversial and is satisfied by a wide variety of two-stage models.
Assume that the observer uses the two-stage model $\mathcal F$ and infers $x \succ y$ given the data set $\ds$.
If she wants to know more about $x$, she may of course collect more data where $x$ is present.
Nonetheless, she could also benefit from a data that does not contain $x$. 
This is because such data may help her infer $y \succ z$, and thus \textit{indirectly} that $x \succ z$ by taking the \textit{transitive closure} of the revealed preference relation. 
Needless to say, it seems natural that the second channel (indirect preferences) could only work if the initial data set tells us something about the lower contour set of $x$; that is indirect revelation may only built on an restrictions in the form of direct revelation. 

\begin{rpproperty}[Standard Transitive Closure (STC)]
\label{axiom:Indirect} 
If a data set does put any restriction on the lower contour set of $x$, then adding more data where $x$ is not available at all does not put any restrictions on the lower contour set of $x$; that is,

$$\text{if} \; \mathcal{R}^{\downarrow} \big(x | (\gB_1,\C_1) \big) = \emptyset, \; \text{and}\; x \notin \bigcup \gB_2, \; \text{then}\; \mathcal{R}^{\downarrow} \big(x |  (\gB_1,\C_1) \cup (\gB_2,\C_2) \big) = \emptyset.$$

\end{rpproperty}

\noindent
Note that if a data set does not put any \textbf{or}-restrictions on the set of inferiors to an alternative ($\mathcal{R}^{\downarrow} (x|\ds) = \emptyset$), then the lower contour set of $x$ will be empty. ($L(x|\ds) = \emptyset$) The reverse is obviously not true. An incipient intuitive inquiry about the principle above hence may be that whether it could be stronger. That is, doesn't it make sense that the extra data where  $x$ is not present should not help identifying an alternative inferior to $x$ \textit{also} in the case where the initial data did not reveal anything inferior to it?-- that is, can we replace $\mathcal{R}^{\downarrow}$ with $L$ in the expression of this principle?

This turns out to be false. To see this take the model where DM shortlists a minimum of two alternatives. Our initial data point is $\C\{x,y,z\} = x$ which tells us that $x \succ ^* y $ \textbf{or} $z \succ ^* t$. Note that given this data we cannot pin down anything inferior to $x$ yet since there is a representation in which $x$ is ranked below $y$ and one where it is below $z$; i.e., $L(x) = \emptyset$. However, we can collect extra data where $x$ is not present, namely $\C\{y,z\} = y$ which indicates $y \succ^* z$. Now there are two possibilities. If $x$ is better than $y$ based on the initial data, then transitivity tells use that it is also better than $z$. In the other possibility $x$ is again better than $z$. Thus, $x$ is revealed better than $z$ for sure, even though the extra data did not contain $x$ at all. This example thus once again points to the relevance of the partial RP information produced by a model through \textbf{or}-logic.\footnote{
    Example such as this can also be applied to the two-stage models in the literature. See \cite{de2021bounded}.}

To further buttress that \textbf{STC} is much less less restrictive than \textbf{CoP}, recall that the two models we argued to violate \textbf{CoP} are rational choice and model where DM always considers at least two alternatives.
Both these models satisfy \textbf{STC}.
Indeed it seems hard to think of a ``standard'' two-stage model that violates \textbf{STC}. 
For example, take the model $\cF = \mathfrak{F}$ where there is no restriction on the filter selection.
In this model, we can always rationalized the data by  ``filtering anything but the choice out''.
That is, there is no welfare content in this model, and thus \textbf{STC} is obviously satisfied.
It also can be shown that the two-stage models listed in Section \ref{sec:models} follow \textbf{STC}. Moreover, these models satisfy \textbf{CoP}.

\begin{remark}
    LA, LC, RSM, and TRSM follow \textbf{CoP} and \textbf{STC}.\footnote{
    Let us make a note here that it is not only these four models, but also all the models listed in footnote \ref{fn:othermodels} that follows these two properties.
    }
\end{remark}

\noindent
Therefore, these four models are members of a class of models defined by the two RP principles. 
Recall that any model in this family faces the observational burden due to \textbf{CoP}: to identify an alternative inferior to $x$, we would require that (i) $x$ to be chosen over an alterative and (ii) the removal of that alternative to also alter choice is some instance. 
It turns out indeed that, when combined with \textbf{STC}, the observational burden becomes more more severe: \textit{instances where (i) and (ii) occur need to be closely related}. 

    

\begin{thm}\label{thm:idencop}
    Assume $\cF$ follows \textbf{CoP} and \textbf{STC}. If $x$ is not involved in a WARP violation -- that is, if there does not exit $T,S,Q \in \gB$ with $x \in T\cap S\cap Q$, and $Q \cap (S\setminus \{x\}) \neq \emptyset$ or $Q \cap (T\setminus \{x\}) \neq \emptyset$, such that 
    \begin{enumerate}
        \item[i.] $(T,S)$ is a WARP violation, 
        \item[ii.] $\C(Q) = x$,

        \end{enumerate}
    
    \noindent then it is not revealed preferred to any other alternative; that is, $L(x|\ds) = \emptyset$.
\end{thm}

\begin{figure}[!h]\caption{Observational Burden beyond \textbf{CoP}}\label{fig:ersm}

        \centering

        \minipage{0.3\textwidth}\subcaption{}\label{fig:CoC}
	\begin{framed}
        \includegraphics[width=\linewidth]{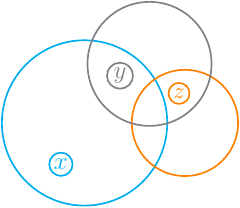}    
        \end{framed}
        \endminipage\hspace{2cm}   
        \minipage{0.3\textwidth}\subcaption{}\label{fig:CoCxSARP}
	\begin{framed}
	\includegraphics[width=\linewidth]{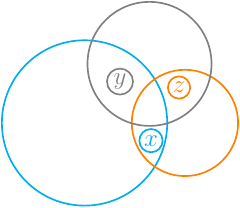}
	\end{framed}
	\endminipage\hfill \\ {\color{white}lkdafjkdjf}\\
        \minipage{0.3\textwidth}
	\begin{framed}
	\includegraphics[width=\linewidth]{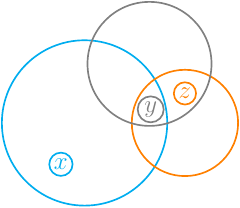}
	\end{framed}
        \subcaption{}\label{fig:CoCxWARP}
	\endminipage

\end{figure} 

\noindent
Theorem \ref{thm:idencop} shows that for an alternative to have non-empty lower contour set it is no sufficient to be chosen over a pivot and it is further required to be involved in a WARP violation.
Figure \ref{fig:Chosen} helps us to see this further requirement in three layers.
First note that being chosen over pivots does not imply violations of rationality.
An instance is illustrated on Figure \ref{fig:CoC}, where $x$ is chosen over the pivot $y$, while there are no violations of rational choice.
Theorem \ref{thm:idencop} implies that nothing can be inferred about the lower contour set of $x$.
More generally, if there are no violations of rationality in a data set, then the same reasoning holds for \textit{any} alternative.
Theorem \ref{thm:idencop} thus implies rather a known feature of these models: \textit{if the data set is consistent with standard rationality (has no SARP violations), then there is no preferences revealed.}
   
    \begin{cor}\label{cor:rational-data}
        Assume $\cF$ satisfies \textbf{CoP} and \textbf{STC}. If that data set is rational -- i.e., no SARP violations -- then $\cF$ is welfare-irrelevant; that is, $\succ^* = \emptyset$.\footnote{While we can use Theorem \ref{thm:idencop}, another simple argument is useful. It is easy to check that a strict linear order satisfies the filter conditions of all these models. Thus, when fed a rational pattern we can attribute the pattern completely to the filter -- anything but the choice is filtered out -- and thus use any preference in the second stage.}
    \end{cor}

\noindent
While known, Corollary \ref{cor:rational-data} may appear paradoxical.
\textit{Identification in these models is linked to the irrational patterns they are intended to explain!} 
Importantly, let us not that this is not an innate feature of two-stage models.
Take, once again, the example of a DM who always considers at least two alternatives.
Abstract away from any consistency conditions on filter formation for a moment. 
We are able to directly infer her preferences in doubletons.
In addition, when the decision problem has more than two alternatives, we would get partial information in the form of the \textbf{or}-logic which, in combination with direct information, could further enrich identification. 
This we can do \textit{whether or not} the data set exhibits violations of rationality. 

The second layer of observational burden implied by Theorem \ref{thm:idencop} is illustrated in Figure \ref{fig:CoCxSARP}.
Here $x$ is chosen over a pivot $y$.
No there is a violation of rational choice: $x$ is chosen over $y$, which, in turn, is chosen over $z$, which is then chosen over $x$.
This a pure SARP violation though, and, hence, $x$ is not involved in a WARP violation.
According to Theorem \ref{thm:idencop} we are yet again unable to say anything about the lower contour set of $x$.

\begin{cor}\label{cor:strictsarp}
    Assume $\cF$ satisfies \textbf{CoP} and \textbf{STC}. If the set of violations in the data set only consists of pure SARP violations, then $\cF$ is welfare-irrelevant; that is, $\succ^* = \emptyset$.
    
\end{cor}

\noindent
It is important to note that Corollary \ref{cor:strictsarp} is only relevant if the data set is incomplete. 
It is a well-known (also straightforward to observe) that if a data set in hand is complete, then any SARP violation in a section of the data will inevitably create a WARP violation somewhere else. 

Finally, Figure \ref{fig:CoCxWARP} presents a case where $x$ is chosen over a pivot, and that there is a case of WARP violation (between the choice problems in gray and orange). 
However, $x$ is not present in the intersection of this violation, and thus, not involved in it.
Once again Theorem \ref{thm:idencop} implies that the lower contour set of $x$ is an empty set.


\subsection{Choice over Choice}
Recall that there are two types of pivots: (i) the chosen ones, and (ii) the unchosen ones the removal of which alters choice.
While \textbf{CoP} is a necessary condition for all four models listed we listed before, it turns out that three of them -- LA, LC, and RSM -- impose an even tighter necessary condition for preference identification.
For these models being chosen over a Type 2 pivot does not allow to draw any inference about the non-emptiness of the lower contour sets.

\begin{rpproperty} [Choice over Choice (CoC)]
\label{axiom:Direct} 
If $x$ is not chosen over a chosen alternative in a data set, then there are no restrictions on the lower contour set of $x$ given that data set; that is,\\

$\forall \ds, \forall x$ if $\not\exists B\in \gB, y\in \C(\gB)$ such that $x= \C(B)$ and $y \in B$, then $\mathcal{R}^{\downarrow}(x |\ds) = \emptyset$. 

\end{rpproperty}

\begin{remark}
    LA, LC, and RSM follow \textbf{CoC}.
\end{remark}

\noindent
Imposing a tighter necessary condition on revealed preferences through \textbf{CoC}, when combined with \textbf{STC}, further exacerbates the observational burden established in Theorem \ref{thm:idencoc}.
In order to make sure that a given alternative has a non-empty lower contour set, the alternative needs to itself \textit{cause} a WARP violation (or be \textit{directly} involved in a WARP violation).

    

\begin{thm}
\label{thm:idencoc} 
Assume $\cF$ follows \textbf{CoC} and \textbf{STC}. If $x$ is not directly involved in a WARP violation in $\ds$ -- that is, if there does not exist $S,T \in \gB$ such that $x = \C(S)$, $S\cap T \ni x,\C(T)$, and $x \neq \C(T)$ -- then $x$ is not revealed preferred to any alternatives; that is, $L(x | \ds) = \emptyset$.


\end{thm}

\begin{figure}[!h]\caption{Necessary Conditions for Pointwise Welfare-Relevance}\label{fig:ersm}	\label{fig:cocvscop}
        \centering

        \minipage{0.30\textwidth}
	\begin{framed}
        \includegraphics[width=\linewidth]{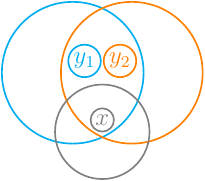}    
        \end{framed}
        \subcaption{\textbf{CoP}: involvement in a WARP violation}\label{fig:indinv}
        \endminipage\hspace{2cm}   
        \minipage{0.30\textwidth}
	\begin{framed}
	\includegraphics[width=\linewidth]{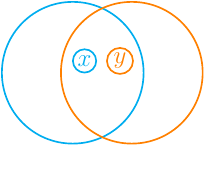}
	\end{framed}
	\subcaption{\textbf{CoC}: direct involvement in a WARP violation}\label{fig:dirinv} 
        \endminipage\hfill

\end{figure} 


\noindent
Note that if $x$ is directly involved in a WARP violation, then it is involved in a WARP violation. 
Figure \ref{fig:cocvscop} visually contrasts these two notions.
Under \textbf{CoC}, in the best case, we can make inference about non-emptiness of lower contour sets for the points that are directly involved in a WARP violation; i.e., two points in total.
Under \textbf{CoP} however a single WARP violation could contribute to identification for more alternatives as long as they are chosen in the intersection of this violation.
Hence, an important takeaway from Theorems \ref{thm:idencop} and \ref{thm:idencoc} is that TRSM has a theoretical potential to perform better than the other three models.


\textbf{CoC} will have a \textit{stronger} welfare-irrelevance implication as long as pure SARP violations are concerned: any alternative that is only involved in pure SARP violation -- that is the choice cycles it is a part of are always of an order higher than two --  will have nothing revealed inferior to it, despite the data set exhibiting WARP violations. This is the content of the next corollary with which we conclude this section.  
\begin{cor}
    Assume that $\cF$ satisfies \textbf{CoC} and \textbf{STC}. If $x$ is only involved in a pure SARP violation, then $L(x|\ds) = \emptyset$.
\end{cor}


\section{Experimental Illustration}
\label{sec:Experiment}

We use the data from \cite{inoue2018limited}.
Let us start by presenting the essential details of the experiment.
Subjects are asked to choose between a bundle of intertemporal installments.
Each bundle consists of three installments: in 1 month, in 3 months, and in 5 months.
Each renumeration bundle consists of 2400 Japanese yen (about \$16), split into three installments.
The universal set consists of 10 alternatives presented in Table \ref{tab:Alternatives}.

\begin{table}[ht]
    \centering
    \resizebox{1\textwidth}{!}{
    \begin{tabular}{l|cccccccccc}
         &  $x_1$ & $x_2$ & $x_3$ & $x_4$ & $x_5$ & $x_6$ & $x_7$ & $x_8$ & $x_9$ & $x_{10}$ \\ \hline
    in 1 month & 450 &  800 & 1150  & 450 &  450 & 800 & 850 & 1200 & 1550 & 500 \\
    in 3 months & 800 & 800 & 800 & 450 & 1500 & 1150 & 0 & 0 & 0 & 0 \\
    in 5 months & 1150 & 800 & 450 & 1500 & 450 & 450 & 1550 & 1200 & 850 & 1900 \\ \hline
    \end{tabular}}
    \caption{Universal space of alternatives. Payments are in Japanese yen.}
    \label{tab:Alternatives}
\end{table}

\begin{figure}[htb]
    \centering
    \includegraphics[width = 1\linewidth]{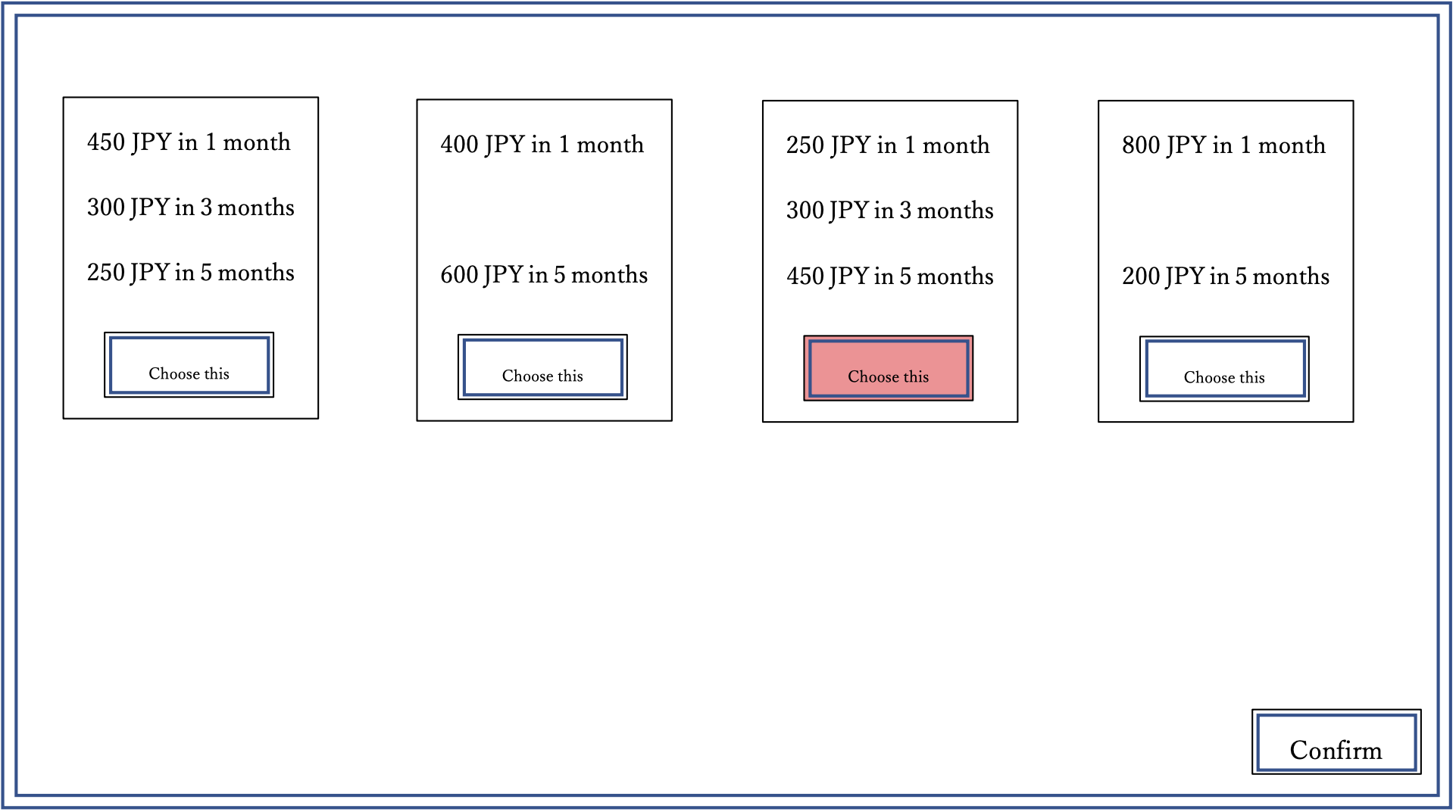}
    \caption{Experimental Interface}
    \label{fig:Interface}
\end{figure}

\noindent
Each subject was presented with 20 different (but same across subjects) choice problems to choose from.
Problems contain two to eight alternatives.\footnote{The composition of the problem sizes is as follows: 6 problems of size two, 2 problems of size 3, 3 problems of size 4, 2 problems of size 5, 3 problems of size 6, 3 problems of size 7, and 1 problem of size 8.}
Figure \ref{fig:Interface} presents a typical experimental interface.
A total of 113 subjects (students of Waseda University, Japan) participated in the experiment that was run in 4 sessions.

The remainder of this section is organized as follows.
We first use the data from the experiment to evaluate the overall possibility of an alternative having non-empty lower contour sets using the results of Theorems \ref{thm:idencop} and \ref{thm:idencoc}; that is, abstracting away from a particular model.
This practice is interesting to get an overall idea about the power of identification, especially because in many cases the observer may believe that the right models of choice is shortlist and choose, but may not know which model is the right one.
It however comes with the cost that detailed information about revealed preferences (what is better than what) can not be given without specifying the model.
Hence, we then zoom in to a given model to actually compute the density of the revealed preference relation.

\subsection{General Welfare-Relevance: Theorems \ref{thm:idencop} and \ref{thm:idencoc}}
Recall that Theorems \ref{thm:idencop} and \ref{thm:idencoc} can help us determine the number of alternatives with non-empty lower contour sets, that is number of alternatives that are revealed preferred to some other.
To put things in perspective, we use the equivalent indicator for the benchmark model of rational choice where it will be equal to the \textit{total number of distinct chosen alternatives}, since any chosen alternative is revealed preferred to another in that model.\footnote{Note that by definition the problems always contain at least two alternatives.}
Next, we have the family of two-stage model that follow \textbf{CoP} and \textbf{STC}.
By Theorem \ref{thm:idencop}, the indicator here is the number of those alternatives that are involved in a WARP violation.
Finally, under Theorem \ref{thm:idencoc}, the indicator for the two-stage models following \textbf{CoC} and \textbf{STC} is the number of alternatives that are directly involved in a WARP violation.

\begin{figure}[p]
\centering
\begin{subfigure}{\textwidth}
        \centering
\begin{tikzpicture}
      \begin{axis}[
      width  = .7*\textwidth,
      height = 7cm,
      major x tick style = transparent,
      ybar=2*\pgflinewidth,
      bar width=20pt,
      ymajorgrids = true,
      xtick = data,
      enlarge x limits=0.1,
      ymin=0,
      ymax=1,
      legend cell align=left,
      legend style={at={(0.5,-0.12)}, anchor = north},
  ]
      \addplot[style={fill=white},error bars/.cd, y dir=both, y explicit]
          coordinates {
    (     0,         0)
    (1.0000,         0)
    (2.0000,         0)
    (3.0000,    0.0088)
    (4.0000,    0.0973)
    (5.0000,    0.5133)
    (6.0000,    0.2389)
    (7.0000,    0.1239)
    (8.0000,    0.0088)
    (9.0000,    0.0088)
   (10.0000,         0)
          };

  \end{axis}

  \end{tikzpicture}
  \caption{Number of distinct chosen alternatives}
  \label{fig:Chosen}
\end{subfigure}

\begin{subfigure}{\textwidth}
        \centering
\begin{tikzpicture}
      \begin{axis}[
      width  = .7*\textwidth,
      height = 7cm,
      major x tick style = transparent,
      ybar=2*\pgflinewidth,
      bar width=20pt,
      ymajorgrids = true,
      xtick = data,
      enlarge x limits=0.1,
      ymin=0,
      ymax=1,
      legend cell align=left,
      legend style={at={(0.5,-0.12)}, anchor = north},
  ]
      \addplot[style={fill=white},error bars/.cd, y dir=both, y explicit]
          coordinates {
    (     0,    0.3451)
    (1.0000,         0)
    (2.0000,    0.0796)
    (3.0000,    0.1593)
    (4.0000,    0.1681)
    (5.0000,    0.1416)
    (6.0000,    0.0973)
    (7.0000,    0.0088)
    (8.0000,    0.0)
    (9.0000,    0.0)
   (10.0000,         0)
          };

  \end{axis}

  \end{tikzpicture}
  \caption{Number of distinct alternatives  involved in WARP violations}
  \label{fig:Pivots}
\end{subfigure}

\begin{subfigure}{\textwidth}
        \centering
\begin{tikzpicture}
      \begin{axis}[
      width  = .7*\textwidth,
      height = 7cm,
      major x tick style = transparent,
      ybar=2*\pgflinewidth,
      bar width=20pt,
      ymajorgrids = true,
      xtick = data,
      enlarge x limits=0.1,
      ymin=0,
      ymax=1,
      legend cell align=left,
      legend style={at={(0.5,-0.12)}, anchor = north},
  ]
      \addplot[style={fill=white},error bars/.cd, y dir=both, y explicit]
          coordinates {
    (     0,    0.3451)
    (1.0000,         0)
    (2.0000,    0.2920)
    (3.0000,    0.1239)
    (4.0000,    0.1239)
    (5.0000,    0.0796)
    (6.0000,    0.0354)
    (7.0000,    0.0)
    (8.0000,    0.0)
    (9.0000,    0.0)
   (10.0000,         0)
          };

  \end{axis}

  \end{tikzpicture}
  \caption{Number of distinct alternatives directly involved in WARP violations}
  \label{fig:Reversals}
\end{subfigure}
\caption{Welfare-relevance analysis.}
\label{fig:Welfare}
\end{figure}
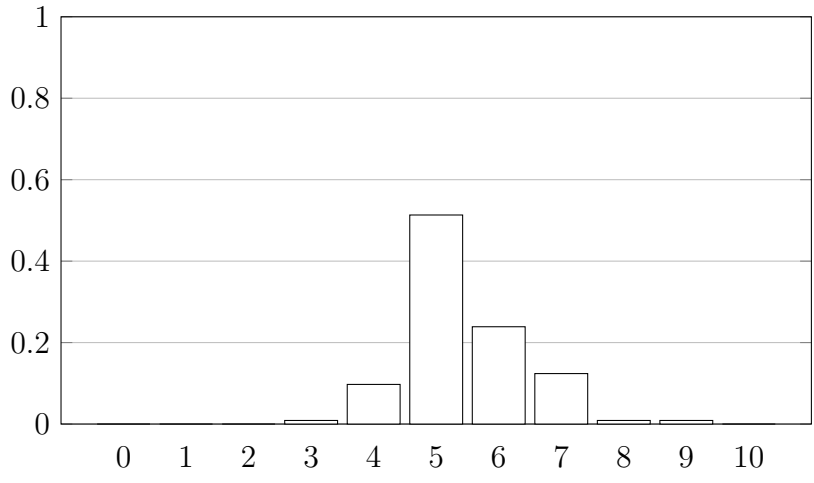
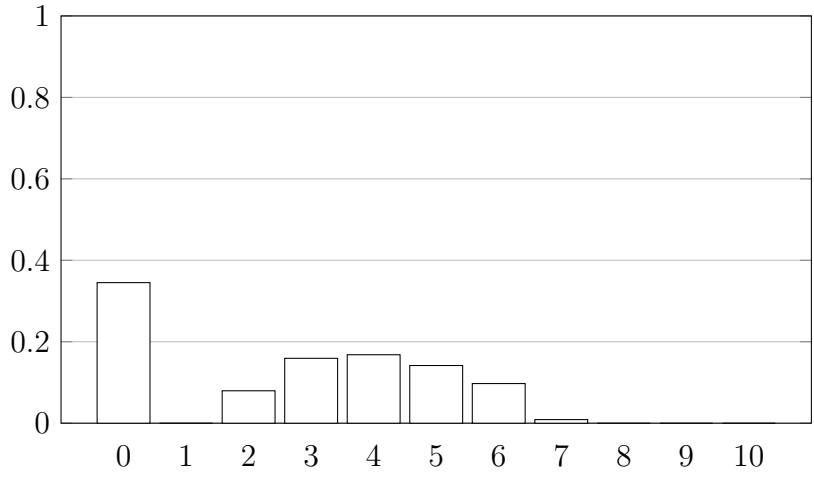
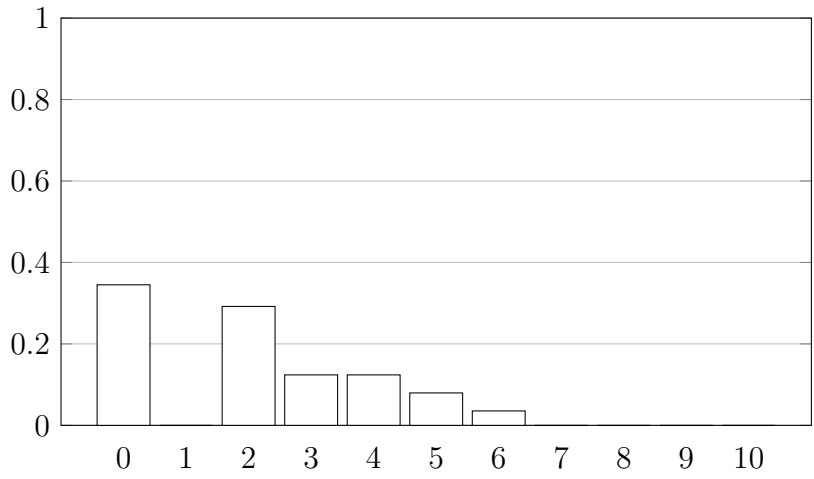

Figure \ref{fig:Welfare} presents the distributions of these three latter indicators. 
Top panel (Figure \ref{fig:Chosen}) shows the distribution of the number of distinct chosen alternatives.
The mean and median here is about 5, with about 40\% of subjects choosing six and more distinct alternatives.
This is notable given that the total number of distinct alternatives is ten.\footnote{We should note that each 10 alternatives are available is some problem, and thus a total of 10 distinct chosen alternative could in principle be observed.}

Middle panel (Figure \ref{fig:Pivots}) presents the number of distinct alternatives involved in WARP violations.
First of all there is mass at zero which corresponds to 34\% of subjects who are rational and thus do not exhibit any WARP violations.
The median number of alternatives with non-empty lower contour set in this case is 2.
That is already less than half than the corresponding number under the assumption of rational choice.
Removing the rational patters however, the distribution will still shift to the left when compared to to the distribution of the distinct chosen alternatives.
The median for the distribution after removing the subjects who look rational is 4.
Thus, in the pool of subjects who violate WARP there are on average about 4 alternatives involved in WARP violations.
This translate to a significant but perhaps not dramatic loss in potential welfare-relevance for a models like TRSM that follow CoP but not CoC.

Finally, bottom panel (Figure \ref{fig:Reversals}) presents the number of distinct alternatives directly involved in WARP violation.
There is again mass at zero, corresponding to 34\% of subjects who are rational and thus do not exhibit any WARP violations.
The median number of alternatives with non-empty lower contour set in this case is 2.
That is already less than half than the corresponding number under the assumption of rational choice.
However, even after controlling for the subjects who look rational we can see that distribution is shifted more to the left compared to the distribution in Panel \ref{fig:Pivots}.
The median for this (updated) distribution is 3 with mean being about the same. 
Thus, even for the subjects who violate WARP there are on average about 3 alternatives directly involved in a WARP violation.
This number is less than a half of the potential inference comparing to rational choice, and about a half compared to those that follow \textbf{CoP} instead.
Therefore, we observe a rather drastic loss in potential welfare-relevance for models like LA, LC, and RSM (that follow \textbf{CoC)} when compared to rational choice.

\subsection{Welfare-Relevance: LA, LC, RSM, and TRSM}
As discussed in the previous section, these models follow \textbf{CoP} (and \textbf{CoC}).
We also intuitively argued why this RP principles  impose a strong restriction on welfare-analysis.
Here, we intend to quantify the extent of this restriction.
In doing so, we make use of the model discussed several times previously: the model that only is concerned with the size of the consideration sets, and not with any consistency conditions.
We denote to this model by NC$_{\text{k}}$ (NC for No-Consistency) where $k$ is the minimum size of the consideration set. 
Note that, for NC$_{\text{1}}$ is a model where \textit{anything goes through}; i.e., DM could be interpreted as if she filters anything but the chosen alternative out.
Thus this model yet follows \textbf{CoP}.
Importantly, on the other hand NC$_{\text{2}}$ \textit{violates} \textbf{CoP} (and thus \textbf{CoC}).
Hence, is a good candidate to quantify the restrictive nature of these properties. 
We then amend this condition to the aforementioned models in order to see how much extra relevance can be brought to a particular theory by a single increment on $k$.
Hence, for each of the four models and NC$_{\text{k}}$ we have two versions: (1) $k=1$ that is original theory, and (2) $k=2$ that is the amended theory assuming that subject makes at least one comparison before choosing.

\paragraph{Descriptive Power}
We start our analysis by evaluating the descriptive power of these models (both original and amended).
That is, we would like to observe to which extent the observed data is consistent with a given model.
The basic indicator here would be the pass rate: the share of subjects who pass the revealed preference test for the corresponding theory (and thus deemed as consistent).
However, different theories would allow for different scope of permissible behaviors.
Thus, the less restriction a model imposes (for example NC$_k$, LA, LC relative to RSM and TRSM that are nested in LC), the higher its pass rate would be.
To correct for this disparity, we also consider the Predictive Success Index (PSI) introduced by \cite{selten1991properties,beatty2011demanding}.
The goal behind PSI is to correct for potential \emph{false positives}.
In order to do so, we generate 1000 subjects who take random decisions at every round.
A false positive then is such a random behavior whenever it is consistent with a given model.
Normalizing the pass rates, PSI subtracts the pass rates of random subjects from that of real subjects.
This index might take values between $-1$ and $1$, where PSI$=-1$ means that every random subject is consistent with the theory while none of the real ones are.
On the other hand PSI$=1$ implies that while every real subject has passed the test, and none of the random ones did.

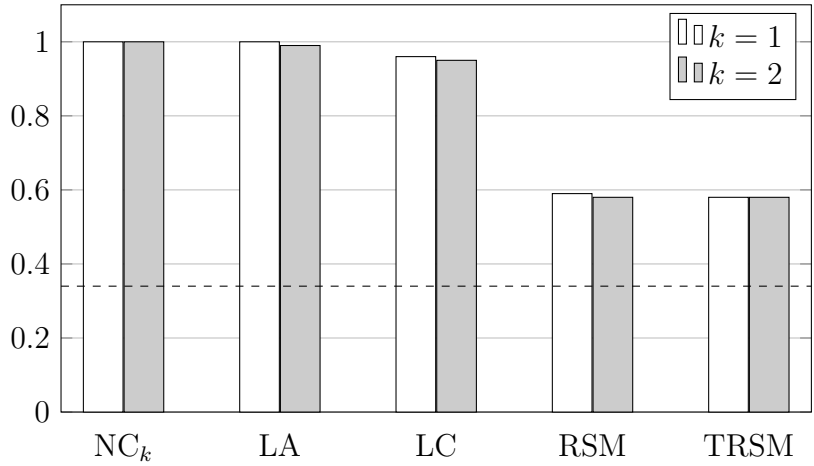
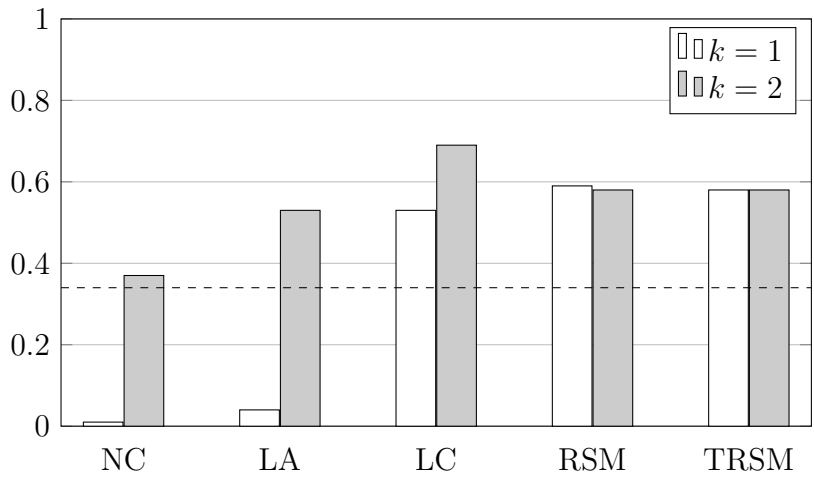
\begin{figure}[p]
\centering
\begin{subfigure}{\textwidth}
\centering
\begin{tikzpicture}
      \begin{axis}[
      width  = .7*\textwidth,
      height = 7cm,
      major x tick style = transparent,
      ybar=\pgflinewidth,
      bar width=15pt,
      ymajorgrids = true,
      symbolic x coords={NC$_k$, LA, LC,  RSM, TRSM},
      xtick = data,
      scaled y ticks = false,
      enlarge x limits=0.1,
      ymin=0,
      legend cell align=left,
      legend style={anchor = north east},
  ]
      \addplot[style={fill=white},error bars/.cd, y dir=both, y explicit]
          coordinates {
          (NC$_k$, 1)
          (LA, 1)
          (LC, .96)
          (RSM, .59)
          (TRSM, .58)
          };
        \addlegendentry{$k=1$}

        \addplot[style={fill=black!20},error bars/.cd, y dir=both, y explicit]
          coordinates {
          (NC$_k$, 1)
          (LA, .99)
          (LC, .95)
          (RSM, .58)
          (TRSM, .58)
          };
        \addlegendentry{$k=2$}

        \path (axis cs:{[normalized]1},.34) coordinate (aux);

  \end{axis}
  \draw[dashed] (current axis.west|-aux) -- (current axis.east|-aux);

  \end{tikzpicture}
  \caption{Pass Rate}
  \label{fig:PassRate}
  \end{subfigure}

\begin{subfigure}{\textwidth}
\centering
\begin{tikzpicture}
      \begin{axis}[
      width  = .7*\textwidth,
      height = 7cm,
      major x tick style = transparent,
      ybar=\pgflinewidth,
      bar width=15pt,
      ymajorgrids = true,
      symbolic x coords={NC, LA, LC, RSM, TRSM},
      xtick = data,
      scaled y ticks = false,
      enlarge x limits=0.1,
      ymin=0, ymax = 1,
      legend cell align=left,
      legend style={anchor = north east},
  ]
      \addplot[style={fill=white},error bars/.cd, y dir=both, y explicit]
          coordinates {
          (NC, 0.01)
          (LA, .04)
          (LC, .53)
          (RSM, .59)
          (TRSM, .58)
          };
        \addlegendentry{$k=1$}

        \addplot[style={fill=black!20},error bars/.cd, y dir=both, y explicit]
          coordinates {
          (NC, .37)
          (LA, .53)
          (LC, .69)
          (RSM, .58)
          (TRSM, .58)
          };
        \addlegendentry{$k=2$}

        \path (axis cs:{[normalized]1},.34) coordinate (aux);

  \end{axis}
  \draw[dashed] (current axis.west|-aux) -- (current axis.east|-aux);

  \end{tikzpicture}
  \caption{PSI}
  \label{fig:PSI}
  \end{subfigure}
  
  \caption{Descriptive power of the theories measured by raw pass rates and the predictive success index (PSI). White bars present the original theory and gray bars present the theories with an additional restriction of making at least one comparison before taking the final decision. Dashed line presents the same measure for rational choice.}
  \label{fig:ConsistencyAnalysis}
\end{figure}

Figure \ref{fig:ConsistencyAnalysis} presents the results of the descriptive power measured as instructed above.
The top panel (Figure \ref{fig:PassRate} shows the raw pass rates for the models of interest.
The dashed line presents the pass rate for the rational choice.
White bars present the pass rates for the model in the original formulation, while gray bars show the pass rates for amend version (at least one comparison in the second stage).
Let us start by noting that all models show the pass rates that are significantly above those for rational choice (33\% for rational choice, while the lowest among the two-stage models is TRSM$\times k=2$ that is 58\%).
This is of course expected as all these two-stage models generalize rational choice.
A couple more interesting observations are the followings.
LC and LA do not create significant drops in pass rates.
More than 90\% of subjects are consistent with either models.
The most interesting observation in this panel is that the pass rate of the original and the amended version of each of these models is almost the same.
In other words, the \textbf{CoP}-violating assumption that a minimum one comparison is made in the second stage has almost \textit{no negative effect on the explanatory power} of these models. 

Figure \ref{fig:PSI} presents the PSI indicator.
The dashed line presents the indicator for the rational choice.
Similar to panel \ref{fig:PassRate}, white bars present the PSI for the original model, while gray bars do the amended version.
Note that since NC$_{\text{1}}$ does not impose any restriction on choice, it is indistinguishable from random choice and thus has a PSI of 0.
Interestingly, increasing the size of the considerations set by one (that is, NC$_{\text{2}}$) raises PSI to the level that is comparable to that of rational choice.
An interpretation is that, while not specifying a particular mechanical channel, such an assumption does take 
For LA we see the picture that is similar to some extent.
The original version of LA seems almost indistinguishable form random choice as indicated in a PSI close to zero.
However, LA's amended version with $k=2$ generates enough restrictions to make the model falsifiable beyond random behavior, with a PSI which is significantly above that of rational choice. 
Finally, for LC, RSM, and TRSM we see that for both versions (original and amended) generate PSIs that are about $.5-.7$; i.e., significantly above that of rational choice ($.33$).
Hence, we can confidently say that these three models explain the data better than rational choice theory.

\begin{figure}[ht]
    \centering
\begin{tikzpicture}
      \begin{axis}[
      width  = .7*\textwidth,
      height = 7cm,
      major x tick style = transparent,
      ybar=2*\pgflinewidth,
      bar width=15pt,
      ymajorgrids = true,
      symbolic x coords={NC, LA, LC, RSM, TRSM},
      xtick = data,
      enlarge x limits=0.1,
      ymin=0,
      ymax=1,
      legend cell align=left,
      legend style={anchor = north east},
  ]
      \addplot[style={fill=white},error bars/.cd, y dir=both, y explicit]
          coordinates {
          (NC, 0)
          (LA, .01)
          (LC, .02)
          (RSM, .01)
          (TRSM, .02)
          };
        \addlegendentry{$k=1$}

        \addplot[style={fill=black!20},error bars/.cd, y dir=both, y explicit]
          coordinates {
          (NC, .17)
          (LA, .20)
          (LC, .20)
          (RSM, .19)
          (TRSM, .21)
          };
        \addlegendentry{$k=2$}

        \path (axis cs:{[normalized]1},.63) coordinate (aux);
  \end{axis}
  \draw[dashed] (current axis.west|-aux) -- (current axis.east|-aux);

  \end{tikzpicture}
  \caption{Average density of the revealed preference relation.
  That is number of the revealed comparisons divided by the number of comparisons to be present in the complete relation (45).
  Dashed line shows the density of the rational revealed preference relation.
  While bars show the results for original theories and gray bars show the results for the theories with the requirement that at least on comparison is performed prior to making the choice.}
  \label{fig:Density}
\end{figure}
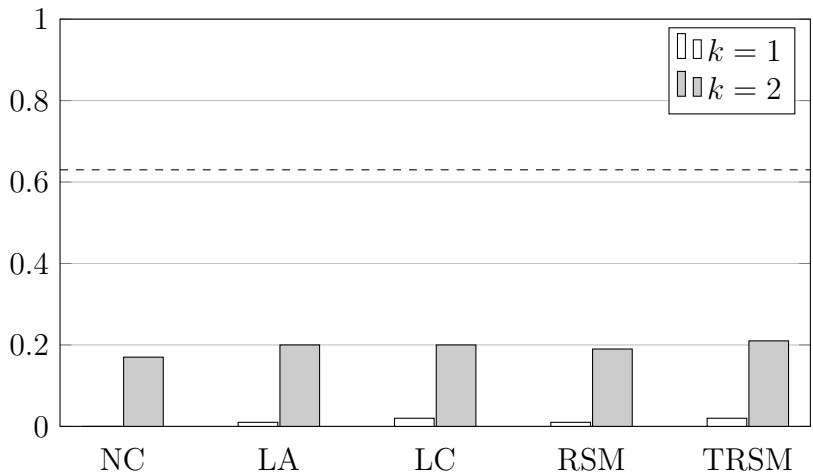

\paragraph{Welfare-Relevance}
Next we proceed to analyze the welfare-relevance of these model by analyzing the density of the revealed preference relation they produce.
By density we mean the number of the revealed comparisons inferred from a model divided by the number of comparisons that would have been present in the complete preference relation (45 in our data since there are 10 alternatives).
Figure \ref{fig:Density} presents the results for the models of interest.
The dashed line presents the density for rational choice.
White bars present the density for the original version, while gray bars correspond to the amended version.
It is important not note that we only compute densities for the subjects who are consistent with given model.

As expected, rational choice is quite efficient in revealing preferences, delivering the density of about 0.62.
However, looking at the original version of the four models we do not observe any notable revelation of preferences.
For these models the density does not exceed .02 which is effectively negligible (less than one comparison per subject passing the model).
Interestingly, there is no significant difference between RSM (following CoC) and TRSM (following CoP).
An interpretation is that, while Theorems \ref{thm:idencop} and \ref{thm:idencoc} open an extra potential channel for welfare-relevance (chosen over Type 2 pivots), TRSM does not appear to be taking advantage of the channel in any effective manner.

Finally, this figure showcases the relative importance of the ad-hoc amendment on the size of the consideration set in welfare-relevance relative to the consistency conditions imposed by each model.
While the original models are equally ineffective, there is significant jump in the density from almost zero to about .20 (0.18-.22 range) upon imposing the amendment.
Interestingly, the intensity of the jump seems to be the same across the models, including NC.
Hence, the major source of identification brought to these models appears to be the amendment not their consistency conditions.
This practice indicates that the specific models we discussed seem to rank quite low in term of welfare-relevance in the class of models that follow these RP principles.

\section{Conclusion}
In a two-stage model DM first makes a shortlist from a given choice problem and then applies her preferences to make a choice.
Following the seminal paper of \cite{manzini2007sequentially}, many models have been proposed, differing on what the right formulation
of the first stage should be.
Prominent examples include but not limited to \cite{masatlioglu2012revealed}, \cite{lleras2017more}, and \cite{au2011sequentially}.
We show that all these models run into serious identification issues by quantifying the density of preference relations revealed.
In the best case one can only infer 2\% of all possible comparisons, while rational choice allows revealing 63\% of all possible comparisons.

We theoretically trace the underlying reasons for this ``welfare-irrelevance''.
A key notion is \emph{pivots} -- alternatives removal of which leads to changing the choice.
These models, along with their counterparts in the literature, all operate on revealed preference principle that we refer to as \textit{choice over pivots}: \textbf{CoP}: \textit{one can not identify the set of inferiors to 
an alternative that is not chosen over a pivot}.
Such an obviously restrictive revealed preference principle is not innate to two-stage models.
It is rather an implication of the fact that these models insist on specifying particular mechanics of the first stage.
Upon making sensible but rather ad-hoc assumptions (such as on the minimum size of the consideration set), two-stage models would violate \textbf{CoP}.

Our experimental analysis confirms \textbf{CoP} to be the source of welfare-irrelevance of these models.
While these models in the original version are almost completely ineffective in revealing preferences (up to 2\%), they all significantly benefit from setting the minimum size of the consideration set equal to two. (about 20\%)
There are two important aspect of our analysis of the data.
First, imposing the minimal restriction on the size of the consideration set has almost no negative effect on the descriptive power irrespective of the model. 
Second, the boost in welfare-relevance is of the same magnitude regardless of the model which is amended by this \textbf{CoP}-violating assumption.
Our analysis overall suggests that the existing two-stage models can benefit from contextually proper but yet ad-hoc assumptions on the way consideration sets are formed a lot more than specifying the ``right'' mechanics of the first stage.


\clearpage
\bibliographystyle{plainnat}
\bibliography{refs}

\begin{thebibliography}{23}
\providecommand{\natexlab}[1]{#1}
\providecommand{\url}[1]{\texttt{#1}}
\expandafter\ifx\csname urlstyle\endcsname\relax
  \providecommand{\doi}[1]{doi: #1}\else
  \providecommand{\doi}{doi: \begingroup \urlstyle{rm}\Url}\fi

\bibitem[Au and Kawai(2011)]{au2011sequentially}
Pak~Hung Au and Keiichi Kawai.
\newblock Sequentially rationalizable choice with transitive rationales.
\newblock \emph{Games and Economic Behavior}, 73\penalty0 (2):\penalty0
  608--614, 2011.

\bibitem[Bajraj and {\"U}lk{\"u}(2015)]{bajraj2015choosing}
Gent Bajraj and Levent {\"U}lk{\"u}.
\newblock Choosing two finalists and the winner.
\newblock \emph{Social Choice and Welfare}, 45:\penalty0 729--744, 2015.

\bibitem[Barseghyan et~al.(2021)Barseghyan, Molinari, and
  Thirkettle]{barseghyan2021discrete}
Levon Barseghyan, Francesca Molinari, and Matthew Thirkettle.
\newblock Discrete choice under risk with limited consideration.
\newblock \emph{American Economic Review}, 111\penalty0 (6):\penalty0
  1972--2006, 2021.

\bibitem[Beatty and Crawford(2011)]{beatty2011demanding}
Timothy~KM Beatty and Ian~A Crawford.
\newblock How demanding is the revealed preference approach to demand?
\newblock \emph{American Economic Review}, 101\penalty0 (6):\penalty0 2782--95,
  2011.

\bibitem[Bernheim and Rangel(2009)]{bernheim2009beyond}
B~Douglas Bernheim and Antonio Rangel.
\newblock Beyond revealed preference: choice-theoretic foundations for
  behavioral welfare economics.
\newblock \emph{The Quarterly Journal of Economics}, 124\penalty0 (1):\penalty0
  51--104, 2009.

\bibitem[Cherepanov et~al.(2013)Cherepanov, Feddersen, and
  Sandroni]{cherepanov2013rationalization}
Vadim Cherepanov, Timothy Feddersen, and Alvaro Sandroni.
\newblock Rationalization.
\newblock \emph{Theoretical Economics}, 8\penalty0 (3):\penalty0 775--800,
  2013.

\bibitem[Dardanoni et~al.(2020)Dardanoni, Manzini, Mariotti, and
  Tyson]{dardanoni2020inferring}
Valentino Dardanoni, Paola Manzini, Marco Mariotti, and Christopher~J Tyson.
\newblock Inferring cognitive heterogeneity from aggregate choices.
\newblock \emph{Econometrica}, 88\penalty0 (3):\penalty0 1269--1296, 2020.

\bibitem[Dardanoni et~al.(2023)Dardanoni, Manzini, Mariotti, Petri, and
  Tyson]{dardanoni2023mixture}
Valentino Dardanoni, Paola Manzini, Marco Mariotti, Henrik Petri, and
  Christopher~J Tyson.
\newblock Mixture choice data: revealing preferences and cognition.
\newblock \emph{Journal of Political Economy}, 131\penalty0 (3):\penalty0
  687--715, 2023.

\bibitem[De~Clippel and Rozen(2021)]{de2021bounded}
Geoffroy De~Clippel and Kareen Rozen.
\newblock Bounded rationality and limited data sets.
\newblock \emph{Theoretical Economics}, 16\penalty0 (2):\penalty0 359--380,
  2021.

\bibitem[Geng(2022)]{geng2022limited}
Sen Geng.
\newblock Limited consideration model with a trigger or a capacity.
\newblock \emph{Journal of Mathematical Economics}, page 102692, 2022.

\bibitem[Geng and {\"O}zbay(2021)]{geng2021shortlisting}
Sen Geng and Erkut~Y {\"O}zbay.
\newblock Shortlisting procedure with a limited capacity.
\newblock \emph{Journal of Mathematical Economics}, 94:\penalty0 102447, 2021.

\bibitem[Horan(2016)]{horan2016simple}
Sean Horan.
\newblock A simple model of two-stage choice.
\newblock \emph{Journal of Economic Theory}, 162:\penalty0 372--406, 2016.

\bibitem[Inoue and Shirai(2022)]{inoue2018limited}
Yuta Inoue and Koji Shirai.
\newblock Limited consideration and limited data: revealed preference tests and
  observable restrictions.
\newblock \emph{Economic Theory}, 2022.

\bibitem[Kimya(2018)]{kimya2018choice}
Mert Kimya.
\newblock Choice, consideration sets, and attribute filters.
\newblock \emph{American Economic Journal: Microeconomics}, 10\penalty0
  (4):\penalty0 223--247, 2018.

\bibitem[Lleras et~al.(2021)Lleras, Masatlioglu, Nakajima, and
  Ozbay]{lleras2021path}
Juan Lleras, Yusufcan Masatlioglu, Daisuke Nakajima, and Erkut Ozbay.
\newblock Path-independent consideration.
\newblock \emph{Games}, 12\penalty0 (1):\penalty0 21, 2021.

\bibitem[Lleras et~al.(2017)Lleras, Masatlioglu, Nakajima, and
  Ozbay]{lleras2017more}
Juan~Sebastian Lleras, Yusufcan Masatlioglu, Daisuke Nakajima, and Erkut~Y
  Ozbay.
\newblock When more is less: Limited consideration.
\newblock \emph{Journal of Economic Theory}, 170:\penalty0 70--85, 2017.

\bibitem[Manzini and Mariotti(2007)]{manzini2007sequentially}
Paola Manzini and Marco Mariotti.
\newblock Sequentially rationalizable choice.
\newblock \emph{American Economic Review}, 97\penalty0 (5):\penalty0
  1824--1839, 2007.

\bibitem[Manzini and Mariotti(2014)]{manzini2014welfare}
Paola Manzini and Marco Mariotti.
\newblock Welfare economics and bounded rationality: the case for model-based
  approaches.
\newblock \emph{Journal of Economic Methodology}, 21\penalty0 (4):\penalty0
  343--360, 2014.

\bibitem[Masatlioglu et~al.(2012)Masatlioglu, Nakajima, and
  Ozbay]{masatlioglu2012revealed}
Yusufcan Masatlioglu, Daisuke Nakajima, and Erkut~Y Ozbay.
\newblock Revealed attention.
\newblock \emph{American Economic Review}, 102\penalty0 (5):\penalty0
  2183--2205, 2012.

\bibitem[Nishimura(2018)]{nishimura2018transitive}
Hiroki Nishimura.
\newblock The transitive core: Inference of welfare from nontransitive
  preference relations.
\newblock \emph{Theoretical Economics}, 13\penalty0 (2):\penalty0 579--606,
  2018.

\bibitem[Selten(1991)]{selten1991properties}
Reinhard Selten.
\newblock Properties of a measure of predictive success.
\newblock \emph{Mathematical social sciences}, 21\penalty0 (2):\penalty0
  153--167, 1991.

\bibitem[Tyson(2013)]{tyson2013behavioral}
Christopher~J Tyson.
\newblock Behavioral implications of shortlisting procedures.
\newblock \emph{Social Choice and Welfare}, 41\penalty0 (4):\penalty0 941--963,
  2013.

\bibitem[Yildiz(2016)]{yildiz2016list}
Kemal Yildiz.
\newblock List-rationalizable choice.
\newblock \emph{Theoretical Economics}, 11\penalty0 (2):\penalty0 587--599,
  2016.

\end{thebibliography}

\clearpage
\appendix

\section{Proofs}

\subsection{Proof of Theorem \ref{thm:idencop}}

Take a given data set $\ds$. Also take an arbitrary $x$. If $x$ is never chosen in the data, then \textbf{CoP} gives us the desired result. So assume that $x$ is chosen somewhere. Assume that $L(x|\ds) \neq \emptyset$. We show that $x$ is involved in a WARP violation. Partition the data set in three parts. (i): $\dsi$ which consists of those problems where $x$ is chosen, (ii) $\dsii$ of those where $x$ is present but not chosen, and (iii) those where $x$ is not present. \textbf{STC} implies that if the observer can not draw any conclusion (infer any restrictions) from the first two partitions about the set of inferiors of a given alternative $x$, then she will not be able to do better by adding the third. That is, she has to infer some restrictions from $\dsi \cup \dsii$. But this would mean that she would need to see $x$ being chosen over a pivot in $\dsi \cup \dsii$.
Let $Q$ be this choice problem and $t$ be the pivot in that problem. Thus, $t \in Q$ and $x = \C(Q)$. 
Given that $t$ is a pivot, there should be problems $S,T\in \gB_2$ (by construction) such that $T=S-t$ and $\C(T)\ne\C(S)$.
Recall that pivot can be either chosen or unchosen alternative.
\\
\textbf{Case 1: $t=\C(T)$.} Since, $T\in \gB_2$ we know that $x\in T$.
Thus, $t,x\in T\cap Q$ and $t=\C(T)\ne\C(Q)=x$.
Thus, $x$ is (directly) involved in WARP violation.
\\
\textbf{Case 2: $t\ne \C(T)$.} Then $\C(S)\in T$ and Thus, we have $\C(T),\C(S)\in S\cap T$ and $\C(S)\ne \C(T)$.
Thus, $(T,S)$ constitute a WARP violation.
Moreover, by construction $x\in T\cap S\cap Q$.
Thus, we conclude that it is involved in a WARP violation.

\subsection{Proof of Theorem \ref{thm:idencoc}}

The proof here is akin to that of Theorem \ref{thm:idencop}. Take a given data set $\ds$. Also take an arbitrary $x$. If $x$ is never chosen in the data, then \textbf{CoC} gives us the desired result. So assume that $x$ is chosen somewhere. Assume that $L(x|\ds) \neq \emptyset$. We show that $x$ is involved in a WARP violation. Partition the data set in three parts. (i): $\dsi$ which consists of those problems where $x$ is chosen, (ii) $\dsii$ of those where $x$ is present but not chosen, and (iii) those where $x$ is not present. \textbf{STC} implies that if the observer can not draw any conclusion (infer any restrictions) from the first two partitions about the set of inferiors of a given alternative $x$, then she will not be able to do better by adding the third. That is, she has to infer some restrictions from $\dsi \cup \dsii$. 
Recall that by construction for every $T\in \gB_1$ the chosen alternative is $x$.
Hence, for $x$ to be chosen over another chosen alternative (as implied by \textbf{CoC}) there should be $y\in T$.
Moreover, we can conclude that $y=\C(S)$ for $S\in \gB_2$.
However, by construction we know that $x\in S$ since $S\in \gB_2$.
Thus, we have $S,T\in \gB_1\cup \gB_2$ such that $x,y\in S\cap T$, while $\C(T)=x$ and $\C(S)=y$.
That is, $x$ needs to be directly involved in a WARP violation.\\

\subsection{Proof of Remarks  1 and 2}

\begin{proof}

\textbf{CoP}. For LA, LC, and RSM we instead show that \textbf{CoC} is satisfied, and thus address the proof of Remark 2 as well. 

\medskip

\textbf{LA.} 
Take a data set that is realizable with LA. Following the argument in \cite{de2021bounded} we know that there exists $\succ$ that satisfies the following set of restrictions:

\begin{equation}
  \begin{aligned}
            & \text{for all} \; S,T \in \gB \; \text{with} \; \C(S) \neq \C(T) \; \text{and} \; \C(S), \C(T) \in S \cap T: \\
            & \C(S) \succ z \; \text{for some} \;  z \in S\setminus T \; \text{or} \; \C(T)  \succ z \; \text{for some} \; z \in T \setminus S
  \end{aligned}
\end{equation}

Now take $x$ that is not chosen over a chosen alternative in the data set. Note that this means there are not restriction of the above form on the lower contour set of $x$ under $\succ^*$.

\medskip

LC. Akin to the line of reasoning before and following a result in \cite{de2021bounded}, this time the set of restrictions are of the form

\begin{equation}
  \C(S) \succ \C (T) \; \text{whenever} \; \C(S) \neq \C(T) \in S \subset T
\end{equation}

Yet, again it follows that if $x$ is not chosen over a choice alternative, there are no restrictions on the lower contour set of $x$ under $\succ^*$.

\medskip

RSM. Let $(\succ_1,\succ_2)$ RSM-rationalize the data set $\ds$. Fix $x$, and Let $A = \{t\neq x: \exists B \in \gB: x,t \in B, x = \C(B)\}$. This would directly imply that $t \succ_1 x$ does not hold for any $t\in A$. Let $\succ^\prime_1 = \succ_1 \cup \{(x,t): t \in A\}$. It follows that $\succ^\prime_1$ is asymmetric. Next, let $\succ_2^\prime$ be a reordering of $\succ_2$ where we move $x$ to from its original rank (wherever) to the bottom of the order. Clearly $(\succ_1^\prime,\succ_2^\prime)$ is a RSM. We show that $(\succ_1^\prime,\succ_2^\prime)$ produces the same choice function on the choice problems. Call this choice function $\C^\prime$. Starting with the $B$ where $x$ is chosen, note that $\succ_1^\prime$ filters anything but $x$ out. Thus $\C(B) = \C^\prime(B)$ Going to the $B$ where $x$ is available, but not chosen, note that a choice in such a problem, say $y$, does not belong to $A$; that $x \succ^\prime y$ does not hold. This implies that neither of the changes implemented has an effect on choice in these problems and thus $\C(B) = \C^\prime(B)$. Lastly, consider $B$ where $x$ is not present. We have only made changes to $x$. Thus, it directly follows that $\C^\prime(B) = \C(B)$. Therefore, $(\succ_1^\prime,\succ_2^\prime)$ is a RSM-rationalization of $\ds$. Since $x =  \argmin\succ_2^\prime$, it follows that no restriction is imposed on the lower contour set of $x$.

\medskip

TRSM. We first prove the following claim.
\medskip

\textbf{Claim}, If an alternative $p$ is not a pivot in the data, then there is always a TRSM-representation $(\succ_1^*,\succ_2)$ where there are no $q$ such that $p \succ_1^* q$. 

To see this claim, take $(\succ_1,\succ_2)$ as a representation where the non-pivot alternative $p$ satisfies $p \succ_1 q \in Q$. Let $\succ_1^\prime$ be a binary relation where we remove all of these latter cases. Note that $\succ_1^\prime$ remains both asymmetric and transitive. Clearly $(\succ_1^\prime,\succ_2)$ generates a TRSM rationalizable data set. We show that this generated data set matches the observed choices. For those problems where $p$ is not available, there is not argument needed. So take those where $p$ is available, and note that, since it is not a pivot it is not the choice in such problems. Next, since we shrunk the filter the consideration set in such sets may expand. So take $x$ that was not considered before the change and now it is considered; that is, $t \succ_1 x$ and there is no comparison between the two in $\succ_1^\prime$. Since $p$ was not a pivot, even if removing it added $x$ to the consideration set before the change the choice would have remained the same. This means the choice alternative before removal is preferred to $x$. Thus, after the change and when $x$ is put back in the consideration set, the choice in the problem remains the same. So we have shown that $(\succ_1^\prime,\succ_2)$ also rationalizes the data set.   
 
Having proved the claim above, the rest of the argument is similar to that of RSM. Let $(\succ_1,\succ_2)$ TRSM-rationalize the data set $\ds$. Based on the claim above, let $(\succ_1^*,\succ_2)$ another TRSM-rationalization where none of the pivots are ranked above any other alternatives according to $\succ_1^*$. Fix $x$, and Let $A = \{t\neq x: \exists B \in \gB: x,t \in B, x = \C(B)\}$. Assume that $t \notin \piv(\gB)$ for all $t \in A$. This would mean that there are also not chosen any where and thus $t \not\succ_1^* x$ for all $t\ in A$. We need to show there are no restriction on the lower contour set of $x$ under $\succ_2$. Let $\succ^\prime_1 = \succ_1^* \cup \{(x,t): t \in A \}$. Since $t \not\succ_1^* x$, it follows that $\succ^\prime_1$ is asymmetric. Since $t$ are not ranked above any other alternative in $\succ^*_1$ then $\succ_1^\prime$ is also transitive. Next, let $\succ_2^\prime$ be a reordering of $\succ_2$ where we move $x$ to from its original rank (wherever) to the bottom of the order. Clearly $(\succ_1^\prime,\succ_2^\prime)$ is a TRSM. We show that $(\succ_1^\prime,\succ_2^\prime)$ produces the same choice function on the choice problems. Call this choice function $\C^\prime$. Starting with the $B$ where $x$ is chosen, note that $\succ_1^\prime$ filters anything but $x$ out. Thus $\C(B) = \C^\prime(B)$ Going to the $B$ where $x$ is available, but not chosen. Say that $y$ is chosen here. Note that we can not have $x \succ_1^\prime y$ since in that case that $x$ is chosen over a pivot; that is $y\notin A$. Thus, neither of the changes we implemented has an effect on choice in this problems and thus $\C(B) = \C^\prime(B)$. Lastly, consider $B$ where $x$ is not present. We have only made changes to $x$. Thus, it directly follows that $\C^\prime(B) = \C(B)$. Therefore, $(\succ_1^\prime,\succ_2^\prime)$ is a TRSM-rationalization of $\ds$. Since $x =  \argmin\succ_2^\prime$, it follows that no restriction is imposed on the lower contour set of $x$.

\bigskip

\noindent\textbf{STC}. For LA and LC, the we can directly conclude this from (1) and (2) and following our previous reasoning using the argument of \cite{de2021bounded}; that is, if $x$ is not present, it can not put new restrictions on the preference relation. 

For RSM the result follows from Corollary 1 and the proof of Theorem 1 in \cite{manzini2007sequentially} where the authors show that given a incomplete data set one can construct $\succ_2$ in the following form:

\[\succ_1 = \{(x,y): \; x = \C\{x,y\}\};\]

Now assume that there are no restriction on the lower-contour set of $x$ given the data set $\dsi$. Hence, adding new data such as $\dsii$ where $x$ is not available will not add any restrictions to enrich $\succ_2$. Thus the result follows.

For TRSM, first note that as argued in \cite{horan2016simple} the models where the second rationale is only transitive (that is the model in \cite{horan2016simple}) is both in terms of identification and axiomatization equivalent to TRSM. Furthermore, Theorem 3 in that paper establishes that $x \succ_2 y $ if and only if there exists a 3-cycle $xyz$ such that $\C\{x,y,z\} \neq z$. Thus adding data where $x$ is not available will not add to $\succ_2$.

\end{proof}

\end{document}